\documentstyle[aps,twocolumn]{revtex}
\begin{document}
\draft
\title{Fluctuations of the energy levels number at the mobility edge.}
\author{Arkadii G. Aronov$^1$ and  Alexander D. Mirlin$^{2\dagger}$}
\address{$^1$ Department of Condensed Matter Physics, Weizmann
Institute of Science, 76100 Rehovot, Israel.}
\address{
$^2$ Institut f\"{u}r Theorie der Kondensierten Materie,
  Universit\"{a}t Karlsruhe, 76128 Karlsruhe, Germany}
\date{\today}
\maketitle
\begin{abstract}
We present a detailed analysis of the behavior of two--level
correlation function $R(s)$ in the disordered sample. We show that
in the vicinity of the Anderson transition as well as in $2d$,
the variance of the number of
levels in an energy band of a width $E$ has a linear behavior
as a function of $E$. This is related with an ``anomalous'' contribution to
the sum rule for $R(s)$, which comes from the ``ballistic'' region of $s$.
\end{abstract}
\narrowtext

The problem of energy level correlation in quantum systems
has been studied intensively after the pioneer work of Wigner \cite{wig}.
The random matrix theory developed
by Wigner and Dyson \cite{dys} is known to describe the statistical properties
of spectra
of various complex systems, such as nuclei, atoms or molecules.
Gor'kov and Eliashberg \cite{GE} put forward an assumption
that  the random matrix theory can be also applied to the problem
of energy level statistics of
a quantum particle subject to  a random potential. This hypothesis was proven
by Efetov \cite{efe}, who studied the two--level correlation function
\begin{equation}
R(s)={1\over \langle\nu\rangle^2}\langle\nu(\epsilon)
\nu(\epsilon+\omega)\rangle-1\ ,
\label{eq1}
\end{equation}
where  $\nu(\epsilon)$ is the density of states at the energy $\epsilon$,
$\langle\ldots\rangle$ denote averaging over realizations of the disorder,
$s=\omega/\Delta$, and $\Delta=(\langle\nu\rangle V)^{-1}$ is the mean level
spacing. As was shown in Ref.\cite{efe}, the  correlator
 $R(s)$ for a disordered mesoscopic metallic sample is described by
the Wigner--Dyson distribution
for $\omega\ll E_c$, where $E_c$ is the Thouless energy; $E_c=D/L^2$, with
$D$ being the diffusion constant and $L$ the sample size. Corrections to the
Wigner--Dyson distribution in this region were studied in a recent paper
\cite{KM}. These corrections are of order $1/g^2$, where
$g=E_c/\Delta\gg 1$ is the dimensionless conductance of the sample.
For $\omega\gg E_c$,
the behavior of the correlation function changes  because the
corresponding time scale $\omega^{-1}$ is smaller than the time $E_c^{-1}$
the particle needs to diffuse through the sample. The smooth
(non-oscillating) part  of the
correlation function in this region  was calculated
in \cite{shkl} by means of the
diffusion perturbation theory.

All these results were obtained for the case of weak disorder
(clean enough sample), when $g\gg 1$. As is well known, when the degree of
disorder increases, the Anderson localization phenomenon takes place. In 3d,
there is a critical value of conductance $g=g_*\sim 1$, which corresponds to
the metal insulator transition. The energy level statistics near this
mobility edge is different from that of a good metal and has attracted a
large interest recently. The first study of this problem was undertaken by
Altshuler et al. \cite{shkl1}, who considered the variance
$\langle\delta N^2(E)\rangle=\langle N^2(E)\rangle-\langle
N(E)\rangle^2$ of the number of levels within a band of a width $E$.
This quantity is related to the
two--level correlator (\ref{eq1}) via
\begin{equation}
\langle\delta N(E)^2\rangle=\int_{-\langle N(E)\rangle}^{\langle N(E)\rangle}
(\langle N(E)\rangle)-|s|) R(s) ds\ ,
\label{eq2}
\end{equation}
or, equivalently,
\begin{equation}
{d \over d\langle N(E)\rangle}
\langle\delta N(E)^2\rangle=\int_{-\langle N(E)\rangle}^{\langle N(E)\rangle}
R(s) ds\ ,
\label{eq2b}
\end{equation}
where $\langle N(E)\rangle=E/\Delta$.
As was shown by Dyson \cite{dys}, the random matrix theory leads to the
logarithmic behavior of the variance $\langle\delta N^2\rangle_{WD}
\sim\ln \langle N\rangle$ for $\langle N\rangle\gg 1$.
In the opposite situation, when all
energy levels are completely uncorrelated (known as the Poisson statistics),
one gets $\langle\delta N^2\rangle_P=\langle N\rangle$. Based on some scaling
considerations as well as on numerical simulations, Altshuler et al.
\cite{shkl1} put forward a conjecture that at the critical point
\begin{equation}
\langle\delta N^2\rangle\simeq\kappa\langle N\rangle
\label{eq2a}
\end {equation}
where $0<\kappa<1$ is certain numerical coefficient.
More recently, Shklovskii et al. \cite{shkl93} introduced the
conception of new universal statistics at the mobility edge.
In Ref.\cite{KLAA}
the correlator $R(s)$ at the mobility edge was studied by means
of perturbation theory combined with scaling assumptions about a form
of the diffusion propagator. It was found that for $s\gg 1$,
\begin{equation}
R(s)\propto s^{-2+\gamma}
\label{eq3}
\end{equation}
where $\gamma<1$ is certain critical index.
Using this result, the relation (\ref{eq2}) and the sum rule
\begin{equation}
\int R(s)ds=0
\label{eq4}
\end{equation}
(which follows from the conservation of the number of energy levels),
the authors of \cite{KLAA} concluded that
\begin{equation}
\langle \delta N^2\rangle\propto N^\gamma
\label{eq5}
\end{equation}
with $\gamma<1$, in contradiction with Ref.\cite{shkl1}.

In this paper, we present a thorough analysis of the behavior of the
correlator $R(s)$ in all range of values of $s$ and for different strength
of the disorder, corresponding to a good metal, a critical region and a
critical point. We show that  the asymptotic behavior (\ref{eq3}) of the
correlator $R(s)$ at $s\gg 1$ does not imply the absence of the
linear term (\ref{eq2a}).
It was assumed in \cite{KLAA} that the universal part of the
correlator $R(s)$ (which is the one surviving in the limit
$E/\Delta=\langle N\rangle=const$, $L\to\infty$) satisfies the sum rule
(\ref{eq4}). We will show however that the sum rule
is fulfilled only if all contributions are taken into account, including
the non-universal contribution of the ``ballistic''
region $\omega\sim 1/\tau$, where
$\tau$ is elastic mean free time.
For this purpose, we will estimate the
contributions to the sum rule from all regions of the variable $s$.

In fact, for the conventional model of a particle in a random potential
the definition (\ref{eq1}) of the correlator and the sum rule relation
(\ref{eq3}) should be modified, when the vicinity of critical point is
considered. The reason is that the Anderson transition point corresponds
to a strong disorder regime $\epsilon\tau\sim 1$, so that the condition
$\omega\sim 1/\tau$ implies $\omega\sim \epsilon$. On the other hand,
the density of states $\nu(\epsilon)$ can be considered as a constant
only for small variations of energy $\omega\ll \epsilon$. This means that
variation of $\nu(E)$ should be taken into account in (\ref{eq1}),
(\ref{eq4}). Besides, the condition $\epsilon\tau\sim 1$ leads to a
breakdown of the perturbation theory, that complicates the analysis of the
``ballistic'' region contribution. To get rid of these problems, we
consider here a different microscopic model which has exactly the same
{\it universal} part of $R(s)$, but whose density of states does not
change within the range of $\omega\sim 1/\tau$. This is so-called
$n$--orbital Wegner model \cite{weg}, which can be considered as a system
of metallic granulae forming a $d$--dimensional lattice, each granula
being coupled to its nearest neighbours. In the limit $n\gg 1$ this model
can be mapped onto
a supersymmetric $\sigma$--model defined on a lattice. The action of this
$\sigma$--model reads \cite{zirn,note}
\begin{equation}
S\{Q\}={\gamma\over 2}\sum_{\langle ij\rangle}\mbox{Str} Q_i Q_j +
\varepsilon \sum_i \mbox{Str} Q_i\Lambda
\label{eq4a}
\end{equation}
where $Q_i=T_i^{-1}\Lambda T_i$ are $4\times 4$ supermatrices defined
on sites $i$ of a $d$--dimensional lattice with a lattice spacing $a$,
$T_i$ belongs to the coset space $U(1,1|2)$, $\Lambda=diag\{1,1,-1,-1\}$
and Str denotes the supertrace. Summation in the first term of
eq.(\ref{eq4a}) goes over pairs $\langle ij\rangle$ of
nearest neighbours. The parameters $\gamma$ and $\varepsilon$
are related to the classical diffusion constant $D$, the density of
states $\nu$ and the frequency $\omega$ as follows:
\begin{equation}
\gamma=\pi\nu D a^{d-2}\ ,\qquad \varepsilon=-i\omega\pi\nu a^d/2\ ,
\label{eq5a}
\end{equation}
To find a detailed description of the model
and relevant mathematical objects, a reader is referred to
Refs.\cite{efe,zirn}.

The two--level correlator can be expressed through a correlation function
of this $\sigma$--model in the following way:
\begin{eqnarray}
R(s)=\mbox{Re}\int \prod_j DQ_j&&\left\{\left[{a^d\over 4V}
\sum_i \mbox{Str} Q_i k\Lambda\right]^2-1\right\} \nonumber\\
&& \times \exp(-S\{Q\})\ ,
\label{eq6}
\end{eqnarray}
where $k=diag\{1,-1,1,-1\}$ and $V$ is the system volume.
An important feature of this lattice
$\sigma$--model is that the corresponding
correlator (\ref{eq6}) satisfies exactly the
sum rule (\ref{eq4})\cite{note9}.
As will become clear below, the continuum version of the
$\sigma$--model does not possess this property: there is a deficiency
of the sum rule for it which is related with the contribution of the
domain of $\omega$ close to an ultraviolet cut-off. This phenomenon is
similar to what is known as quantum anomalies in the quantum field theory
(see e.g. \cite{anom}). To prove the sum rule for the
correlator (\ref{eq6}) on a lattice, we substitute eq.(\ref{eq4a}) into
eq.(\ref{eq6}) and integrate over $\omega$. This gives
\begin{eqnarray}
&&\int_{-\infty}^{\infty}\!\! R(s)ds=\mbox{Re}\int \prod_j DQ_j
\left\{ \left[{a^d\over 4V}\sum_i\mbox{Str}
Q_i k\Lambda\right]^2-1\right\}
\nonumber\\
&&\times\exp\left\{-{\gamma\over 2}\sum_{\langle ij\rangle}\mbox{Str}
Q_i Q_j\right\}\:\delta\left({a^d\over 4V}\sum_i\mbox{Str} Q_i\Lambda
\right)
\label{eq7}
\end{eqnarray}
In the conventional way \cite{efe,zirn}, the $Q$--matrices can be
parametrized by 2 ``eigenvalues'' $1\le\lambda_1<\infty$,
$-1\le \lambda_2\le 1$ and a certain set of ``angular'' variables.
We do not need an explicit form of this parametrization here, but only
the fact that $\mbox{Str}Q\Lambda = 2(\lambda_1-\lambda_2)\ge 0$, so
that $\mbox{Str}Q\Lambda=0$ only when $\lambda_1=\lambda_2=1$. Therefore,
the hyperplane $\sum_i(\lambda_1^{(i)}-\lambda_2^{(i)})=0$ determined
by the $\delta$--function in (\ref{eq7}) touches the actual domain of
integration only at one point: $\lambda_1^{(i)}=\lambda_2^{(i)}=1$ for all
i, that leads to vanishing of the integral \cite{note2}.

Let us stress that in the region $\omega\ll D/a^2\equiv E_c(a)$ the
correlator $R(s)$ is universal, i.e. does not depend on microscopical
details of the model. The region $\omega\sim D/a^2$ plays a role
analogous to the ballistic one $\omega\sim 1/\tau$ in the case of
a usual model of particle in a random potential. Despite the
non-universality of the correlation function $R_{nu}(s)$ in this
``ballistic'' domain, the corresponding integral contribution $I_{nu}$
to the sum
rule is universal, because it determines, according
to eq.(\ref{eq4}), the sum rule deficiency for the universal part
$R_{u}(s)$:
\begin{eqnarray}
&&I_{u}+I_{nu}=0\ ;  \nonumber\\
&&I_u=\int R_u(s) ds\ ;\qquad I_{nu}=\int R_{nu}(s) ds
\label{eq8}
\end{eqnarray}
As will be seen below,
when the system is close to the Anderson transition
the two regions of the variable $s$ dominating the
integrals $I_u$ and $I_{nu}$ respectively are separated
by a parametrically broad domain of $s$ which
gives a negligible contribution to the sum rule. For energy band width $E$
lying in this region, we get from eqs.(\ref{eq2b}), (\ref{eq8})
\begin{equation}
\langle\delta N(E)^2\rangle\simeq I_u\langle N(E)\rangle=
-I_{nu}\langle N(E)\rangle\ ,
\label{eq9}
\end{equation}
i.e. just the linear term (\ref{eq2a}), which was conjectured
in Ref.\cite{shkl1}. In Ref.\cite{KLAA}  the ``anomalous''
contribution $I_{nu}$ was lost, that led to a
conclusion about the absence of this linear term.

We come now to the analysis of the correlator $R(s)$ and of the sum
rule in various situation. We start from the case of a good metal,
then we consider the critical point and the critical region cases.

1. Good metal. Here the following three regions with different behavior
of $R(s)$ can be found:

A. ``Wigner--Dyson'' (WD) region: $\omega\ll E_c$. The correlator
$R(s)$ in this region is given by the random matrix theory result
$R_{WD}(s)$
plus a $1/g^2$ correction \cite{KM}:
\begin{equation}
R(s)\simeq R_{WD}(s)+{a_d\over \pi^2 g^2}\sin^2(\pi s)\ ,
\label{eq9b}
\end{equation}
where $a_d>0$ is certain numerical coefficient. Corresponding contribution
to the sum rule can be estimated as
\begin{equation}
I_A\simeq\int_0^g R(s)\propto +1/g
\label{eq9a}
\end{equation}
In all estimates we omit numerical factors of order
unity, keeping only a sign and a parametric dependence.

B. ``Altshuler--Shklovskii'' (AS) region:
$E_c\ll\omega\ll E_c(a)$. Here $E_c(a)=D/a^2$ is the
Thouless energy at the scale of lattice spacing $a$, which plays a role
of the ultraviolet cut-off for the diffusion theory. We have
\cite{shkl}:
\begin{equation}
R(s)\propto +g^{-d/2} s^{d/2-2}\ ,
\label{eq10}
\end{equation}
and consequently,
\begin{equation}
I_B\simeq\int_g^{D/a^2 \Delta } R(s) ds\propto
g^{-d/2}\left({D\over a^2\Delta}\right)^{{d\over 2}-1}\propto +1/g(a)\ ,
\label{eq11}
\end{equation}
where $g(a)=g(a/L)^{d-2}$ is the conductance at the scale $a$. Note that
for the case of a particle in a random potential, the following substitutions
should be done: $a\to l$; $E_c(a)\to E_c(l)=1/\tau$;
$1/g(a)\to 1/g(l)\propto (\epsilon\tau)^{d-1}$.

C. ``Ballistic'' region: $\omega\gtrsim E_c(a)$. (As was already
noted, this region
is not really ballistic in the lattice $\sigma$--model, but plays exactly
the same role as the ballistic one in the usual case.) To find the
correlator $R(s)$ in this region, we can neglect to the first approximation
the first term in eq.(\ref{eq4a}), that gives
\begin{eqnarray}
&&R(s)\propto - \left({L\over a}\right)^d {1\over s^2}\ , \label{eq12}\\
&&I_C\simeq \int_{D/a^2\Delta}^\infty R(s)ds\propto -1/g(a)
\label{eq13}
\end{eqnarray}
The contribution (\ref{eq9a}) to the sum rule is dominated by the
region $\omega\sim\Delta$, whereas the contributions (\ref{eq11}),
(\ref{eq13}) are dominated by a vicinity of the ultraviolet cut-off
$\omega\sim D/a^2$. In $d=3$ $g\gg g(a)$, and these non-universal
contributions $I_B$, $I_C$ are much larger (by absolute value) than $I_A$,
but they should cancel each other according to the sum rule. The situation
is more interesting in $d=2$, where the contribution (\ref{eq10}),
(\ref{eq11}) of the AS region is absent in view of special analytical
properties of the diffusion propagator \cite{gefen}. In this case $g(a)=g$,
so that the contributions $I_A$ and $I_C$ are of the same order,
in agreement with the sum rule which prescribes their sum to be zero.
We can identify then $I_A$ with $I_u$ and $I_C$ with $I_{nu}$
in eq.(\ref{eq8}), so that $I_u=-I_{nu}\propto 1/g$.
According to eq.(\ref{eq9}), this leads to a linear term for the
variance:
 \begin{equation}
\langle \delta N^2\rangle\propto (1/g) \langle N\rangle\ ;\qquad
E_c\ll E\ll D/a^2\ ,
\label{eq14}
\end{equation}
but the corresponding coefficient is of order $1/g\ll 1$.

With an increase in  disorder strength, the coupling constant $\gamma$
in eq.(\ref{eq4a}) decreases. When it approaches a critical value $\gamma_c$,
corresponding to the metal--insulator transition, the correlation
length $\xi$ becomes large: $\xi\gg a$. Depending on the relation between
$\xi$ and the system size $L$, the following two situations may take place.

2. Metal in critical region: $a\ll\xi\ll L$. In this case $g(a)\simeq g_*$,
but $g\equiv g(L)\gg g_*$, where $g_*$ is the value of conductance in the
critical point \cite{note3}. We find then as much as four different regions of
$s$ for the correlator $R(s)$:

A. WD region: $\omega\ll E_c$. The correlator $R(s)$ and the contribution
to the sum rule for this region are given by the same eqs.(\ref{eq9b}),
(\ref{eq9a}), as for a good metal.

B. AS region: $E_c\ll \omega\ll E_c(\xi)=g_*\Delta_\xi$, where $E_c(\xi)$
and $\Delta_\xi$ are Thouless energy and level spacing for a piece of the
sample with linear size $\xi$. The correlator $R(s)$ is given by
eq.(\ref{eq10}), and we find
\begin{equation}
I_B\simeq\int_g^{g_*\Delta_\xi/\Delta} R(s)ds
\propto g^{-d/2}g_*\left({\Delta_\xi\over\Delta}\right)^{{d\over 2}-1}
\propto +1/g_*\ ,
\end{equation}
where we have used the relations $\Delta_\xi/\Delta=(L/\xi)^d$
and $g/g_*\propto(L/\xi)^{d-2}$.

C. ``Kravtsov--Lerner--Altshuler--Aronov'' (KLAA) region:
$E_c(\xi)\equiv g_*\Delta_\xi\ll\omega\ll g_*\Delta_a\equiv E_c(a)$.
The correlator $R(s)$ in this region was studied in Ref.\cite{AKL2};
the result reads:
\begin{equation}
R(s)\simeq -g_*^{-\gamma}\left({\Delta_\xi/\Delta}\right)^{1-\gamma}
s^{-2+\gamma}\ ;
\label{eq16}
\end{equation}
and consequently
\begin{equation}
I_C\simeq\int_{g_*\Delta_\xi/\Delta}^{g_*\Delta_a/\Delta}R(s)ds
\propto -1/g_*
\label{eq17}
\end{equation}

D. ``Ballistic'' region: $\omega\gtrsim E_c(a)$. Here the equations
(\ref{eq12}), (\ref{eq13}) hold, yielding
\begin{equation}
I_D\propto -1/g_*
\label{eq18}
\end{equation}

The contribution $I_A$ is dominated by the domain $\omega\sim\Delta$,
the contributions $I_B,\: I_C$ by $\omega\sim\Delta_\xi$ and finally
the contribution $I_D$ comes from the ``non-universal'' region
$\omega\sim E_c(a)$. Therefore, this last contribution determines
$I_{nu}$ in eq.(\ref{eq8}), and according to eq.(\ref{eq9}) we find the
linear behavior of the variance
\begin{equation}
\langle \delta N^2\rangle\propto (1/g_*)\langle N\rangle\ ;
\qquad E_c(\xi)\ll E\ll E_c(a)
\label{eq19}
\end{equation}
with the coefficient $1/g_*$, which is of order unity in $3d$.

3. Finally, a system is at the critical point, when $\xi\gg L$. In this case,
$g(L)=g(a)=g_*$. The AS region disappears, and we have,
in full analogy with previous estimates:

A. WD region: $\omega\lesssim E_c=g_*\Delta$.
\begin{equation}
I_A\propto +1/g_*
\label{eq20}
\end{equation}

 B. KLAA region: $g_*\Delta\ll\omega\ll \Delta_a g_*$.
\begin{equation}
 R(s)\propto-g_*^{-\gamma}s^{-2+\gamma}\ ;\qquad
 I_B\propto -1/g_*
\label{eq21}
\end{equation}

C. ``Ballistic'' region. $\omega\gtrsim\Delta_a g_*$.
\begin{equation}
I_C\propto -1/g_*
\label{eq22}
\end{equation}

Again the same conclusion, as for the critical region case, can be drawn:
the ``ballistic'' contribution (\ref{eq22}) can be identified as $I_{nu}$ in
eq.(\ref{eq8}), that gives:
\begin{equation}
\langle \delta N^2\rangle\propto (1/g_*)\langle N\rangle\ ;\qquad
E_c\ll E\ll E_c(a)
\label{eq23}
\end{equation}

This concludes our analysis. In fact, all obtained results can be summarized
in the following relation:
\begin{equation}
\int_{-\omega/\Delta}^{\omega/\Delta} R(s) ds=
-\int_{|s|>\omega/\Delta} R(s) ds\propto 1/g(L_\omega)\ ,
\label{eq24}
\end{equation}
where $L_\omega$ is defined by the condition $E_c(L_\omega)=\omega$, and
$g(L_\omega)$ is the conductance of a piece of the sample of the size
$L_\omega$. Equation (\ref{eq24}) holds for all $\omega$ such that
$L_\omega\le L$. Its physical meaning is quite transparent: a portion
of a state ``expelled'' from the volume with linear size $L_\omega$
is inverse proportional to its conductance.

In conclusion, we have demonstrated that the levels number variance (\ref{eq2})
shows a linear behavior (\ref{eq19}), (\ref{eq23})
in the vicinity of Anderson transition, in agreement with an earlier
prediction \cite{shkl1}. The corresponding coefficient
is proportional to $1/g_*$, with $g_*$ being the critical value of
conductance.  We have
shown that this linear term is related to an  ``anomalous'' contribution
to the sum rule for two--level correlation function, which
originates from ``ballistic'' region of frequencies providing an ultraviolet
cut-off for the diffusion theory.

To avoid possible confusion, let us note that one should distinguish
the asymptotic behavior of $\langle\delta N^2\rangle$ from that of
another frequently studied quantity -- spacing distribution
$P(s)$. The former is by definition the variance of the distribution
function of number of levels in an energy band of given width, whereas
the latter is determined by the far tail of this distribution
function. The asymptotic behavior of $P(s)$ at the mobility edge was
found in Ref.\cite{AKL1} to be $P(s)\sim\exp(-s^{2-\gamma})$. We
believe that the
validity of this result is not affected by the present study.

We are grateful to V.E.Kravtsov for numerous discussions which
stimulated appearence of this work. A.G.A. thanks B.Shapiro for
attracting his attention to the problem.
A.D.M. acknowledges discussions of the results
with Y.Gefen,  F.Izrailev and Y.V.Fyodorov and
 is grateful to the International Centre for Theoretical Physics
in Trieste
and the Institute for Nuclear Theory at the University of Washington
for the kind hospitality and to the Alexander von Humboldt Foundation
for  support.


\begin{references}
\bibitem{note} In fact, Eqs.(\ref{eq4a})--(\ref{eq7}) are written for the
case of $\sigma$--model of unitary symmetry, corresponding to broken
time reversal invariance; all results are however completely analogous for
orthogonal and symplectic symmetry cases.
\bibitem{note3} In $3d$ the value of $g_*$ is of course of order unity.
However, if one consider formally $d=2+\epsilon$ with $\epsilon\ll 1$, then
$g_*\sim 1/\epsilon$ and is therefore parametrically large. That is why
it is reasonable to keep $g_*$ as a parameter in all estimates.
\end{references}
\end{document}